\documentclass[twocolumns]{aa}
\usepackage{graphicx}
\usepackage{amssymb}
\usepackage{txfonts}

\begin{document}

\title{Analysis of the white--light flickering of the
Intermediate Polar V709 Cas with wavelets and Hurst analysis.}

\author{F. Tamburini$^1$, D. De Martino$^2$, A. Bianchini$^1$.}

\institute{$^1$ Department of Astronomy, University of Padova, vicolo
dell' Osservatorio 3, IT-35122, Padova, Italy.\\
$^2$ INAF--Astronomical Observatory of Capodimonte, Moiarello 16,
80131, Naples, Italy.}

\date{Received ; accepted}

\abstract{We characterize the flickering observed in the optical lightcurve of the Intermediate Polar system V709 Cas by determining its position in the $\alpha$-$\Sigma$ as in the Fritz and Bruch (1998) classification scheme. 
$\Sigma$ represents the strength of flickering at a given timescale, while $\alpha$ describes the energy distribution of the flickering at different time scales. Here $\alpha$ is independently obtained with both the wavelets and the Hurst R/S analysis.
The flickering shows self--similarity in the time scale ranging from tens of minutes down to 10 seconds with stochastic persistent memory in time. $\alpha$ and $\Sigma$ appear anticorrelated. 
In the $\alpha$-$\Sigma$ diagram  V709 Cas falls in the region of magnetic systems. Since V709 Cas shows the spin period of the magnetic WD only in the X-ray but not in the optical, we conclude that this method can be used  to characterize CV subtypes especially when their classification is uncertain.

\keywords{ Flickering -- Hurst's analysis -- Accretion -- turbulent
phenomena}}

\authorrunning{F. Tamburini et al.}
\titlerunning{Wavelet and R/S analysis of V709 Cas flickering.}
\maketitle

\section{Introduction}

Cataclysmic Variables (CVs) are binary systems in which a
late--type secondary star fills its Roche lobe and transfers
matter onto a white dwarf (WD) primary.
The WD is surrounded by an accretion disk, unless
its magnetic field is strong enough to partially or totally
control the accretion geometry. 

CVs are classified into three major groups: nova-like (NL)
systems, classical and recurrent novae (CN, RN), and dwarf novae (DN).
In particular, nova systems must host CNO WDs with masses
larger than $\sim 0.6 M_{\odot}$ so that the accreted material can
be cyclically  ejected through nova outbursts (Livio 1992).
CVs can also be classified according to the strength of the magnetic
field of the primary. Thus, we have the non--magnetic systems,
the {\it Intermediate Polar} systems
or DQ Her systems (hereafter IPs), and the {\it Polars} or AM Her systems.
In the non-magnetic systems  accretion onto the WD occurs from
the last stable orbit of the inner disk.
In IP systems the magnetic field  disrupts the inner regions of
the accretion disk within the  so-called Alfv\'en radius and
accretion mainly occurs via accretion curtains onto the magnetic
poles of the WD.
In Polar systems the disk is totally disrupted by the strong
magnetic field of the WD and the accretion stream is directly
conveyed on the WD magnetic poles.
In both Polars and IPs the infalling matter forms  a strong
shock above the magnetic poles of the compact star that mainly
radiate in the X--ray domain
and eventually producing cyclotron radiation in the optical/IR.

At any given orbital period, the luminosities of  NL systems and
of quiescent novae are systematically brighter than those of DN,
suggesting that the former ones are powered by larger mass transfer rates
($\dot{M_2}$) from the secondary and/or have hotter WDs preventing
the onset of the cyclical disk instability phenomena  observed in DN.
An exception may be represented by those CN and NL that are also magnetic
Polars in which the strong
magnetic field of the WD reduces the efficiency of the magnetic
braking mechanism of the binary system and, consequently, the mass
transfer rate and the accretion luminosity (\cite{warner95}).

The lightcurves of Cvs may show a variety of periodic, quasi-periodic and/or erratic  modulations.
Flickering consists of short term variations on time scales from a few seconds to a few tens of minutes, with magnitude variations from a few tenths of magnitude to the magnitude scale.
Flickering  appears as a sequence of overlapping flares and bursts with a random variability in time, sometimes showing self--similarity at different time scales, i.e. the stochastic fluctuations in the detrended lightcurve are similar independently from the chosen time binning.
Warner \& Nather (1971) first tried to associate the
flickering  of U Gem and DN to the hot--spot in the outer rim of the AD although this
\textit{ansatz} was not confirmed in the observations of
some other CVs like  VW Hyi (\cite{warner75, vanam87}), Z Cha,
V2051 Oph, (\cite{warodo87}).
The observations of AE Aqr and YZ Cnc  (Elsworth \& James 1982, 1986; 
James, 1987), instead, suggested that the origin of flickering might
lie in the innermost regions of the  AD or close to the surface of the
accreting WD.
Some  models suggest that the instabilities in the mass accretion  onto
the WD (Bruch 1992, 1992a, 1992b, 1996) are due to MHD plasma turbulence in the
accretion flow. 
Recently, the flickering of CVs was studied in several systematic high speed photometry campaigns (\cite{ww01,ww02,ww02b,ww03,ww03b,w04,w05,ww08}).

Flickering has been observed both in non magnetic and magnetic CVs. Magnetic CVs
usually present flickering in the optical and sometimes in the X--ray region of the 
spectrum.
One general property of flickering is the correlation between its amplitude and the luminosity of the ``quiet primary''. The  ``quiet primary'' is defined as the sum of the luminosity of the WD and the luminosity of that part of the accretion process (disk or flow) which is not involved in the flickering activity (Zamanov \& Bruch, 1998).
Information can be also obtained by comparing the flare rate and the ratio of the maximum and the mean flux observed to that of the quiet primary.
In some objects the flickering involves a large fraction of the total light emitted  eventually dominating the optical light curve.
The modalities of the flickering observed may vary from object to object but some common features have been observed in CVs of the same sub-type.
Bruch et al. (Bruch 1992) proposed a classification method of the flickering showing that different subtypes of CVs tend to occupy specific regions of the $\alpha$-$\Sigma$ plane, where $\alpha$
is the parameter related to the energy distribution of the flickering at different time scales and $\Sigma$ represents the strength of the flickering for a given time scale. 

In this paper we derive the $\alpha$ parameter for the IP system V709 Cas both using the wavelets and the Hurst R/S analysis (Hurst et al. 1965) and discuss the two methods.
While the wavelet analysis of flickering is dependent on the choice of the mother wavelet function, Hurst's R/S analysis bypasses this problem, giving an independent determination of $\alpha$. 
In addition, the value of the Hurst exponent gives the degree of persistence/antipersistence of the stochastic memory of the flickering, depicting the global behavior of the physics of the accretion process.
Differently from the X-ray band, this magnetic CV shows no spin modulation in the optical 
range and it is strongly flickering--dominated, without any precise and stable dominating frequency. 
V709 Cas hence represents an ideal test case for our study.

In Sec. 2 we introduce the mathematical basis for both wavelets and Hurst's analysis and explain the link between these two methods.
In Sec. 3 we characterize the flickering of this IP system by placing it in the $\alpha$, $\Sigma$ diagram. Sect. 4 draws the conclusions.

\section{Flickering properties and Hurst exponent}

In the past, many authors discussed flickering mainly in terms of statistical properties (\cite{mum66, rob73, mb74, zuk71}), Bruch (1992, 1992b, 1996), Bruch \& Gr\"utter (1997).
A further step was made by Fritz \& Bruch (1998) who, following Scargle et al. (1993), introduced the use of wavelet transforms and of the scalegram.
The scalegram is a plot of the logarithm of a reference timescale $t_s$ vs the logarithm of the variance $S(t_s)$, namely a measure of the variances of the wavelet coefficients expressed as a function of the time scale. See also \cite{pw00} for more details.

Flickering is a stochastic process in time whose properties can provide vital information about the driving mechanism behind the accretion process. One of the most important statistical properties of flickering is the measure of the variance of the wavelet coefficients at different time scales
\begin{equation}
S'(t_s)=\frac{2^s}{N}\sum_k c^2_{s,k}
\end{equation}
where $c^2_{s,k}$ are the wavelet coefficients in which $k$ is the time index, $N$ the number of measurements (equivalent to the total number of the wavelet coefficients) and $s$ is the scale index, also called ``octave'', related to the time scale by $t_s=2^s \Delta t$.
$S'(t_s)$ is related with the total energy of the signal calculated at the time scale $t_s$ and summed over all the time bins, then normalized to N and to the integration interval $\Delta t$.
The function $S'(t_s)$ characterizes the statistical variations of the data record in the time interval $[2^s\Delta t, 2^{s+1}\Delta t]$.

Fritz \& Bruch (1998) and Zamanov \& Bruch (1998) showed that the scalegrams of flickering are with a good approximation linear functions of $t_s$ for most of the CVs' lightcurves. This implies the presence of self--similarity with a power--law behavior $S'(t_s)\sim s^\alpha$ within certain time scales.
The stochastic fluctuations in the detrended lightcurve are similar independently from the time binning as described by $1/f^\beta$ noises, where $f$ is the time frequency and $\beta=\alpha/2$.

The linearity of the scalegrams permits a simple parametrization of the flickering properties with two parameters, $\alpha$ and $\Sigma$. $\alpha$ is the inclination of the scalegram with respect to the $x$ axis, that can be then determined with a linear least squares fit to the scalegram points of the $\log S(t_s)$, $\log (t_s)$ plot. 
$\alpha$ shows whether slow or rapid light fluctuations are dominating the stochastic time series and the flickering strength parameter $\Sigma=log S(t_s) \cdot t_s$, given by
\begin{equation}
S(t_s)=S'(t_s)\frac{N \Delta t}{\sum_{s,k} c^2_{s,k}}
\end{equation}
is given by the measure of the variance of the wavelet coefficients for a given time scale.
Some possible correlations between the strength and the  duration of the stochastic light variations can be put in evidence by plotting $\alpha$ $vs$ $\Sigma$ in the so-called $\alpha$-$\Sigma$ parameter space.

An alternative robust statistical tool that can characterize independently from the wavelet approach the stochastic properties of flickering (i.e. the parameter $\alpha$) is Hurst's R/S analysis. This method offers the advantage of being independent from any previous modeling of the problem, like the choice of a mother wavelet.
The approach initially used by Hurst (1951) and Hurst et al. (1965) is based on the rescaled range analysis which consists in estimating the ratio between the range of the variations, R, and the standard deviation S derived from the analysis of all the sub-intervals of data for each equal partitioning of the full data record.
Mandelbrot \& van Ness (1968) and Mandelbrot \& Wallis (1969) linked this method to a particular class of self--similar random processes, named Fractional Brownian Motions (FBMs). It was also shown that also short--run statistical dependence in pseudo-random sequences can be described with this technique (\cite{gam97}).
Hurst found that natural phenomena follow the empirical power law
\begin{equation}
R/S= (\epsilon N)^H
\label{eqh}
\end{equation}
where $\epsilon$ is a constant, $N$ the number of data present in each sub-interval,
and $H$ is the so-called Hurst's exponent.
In particular, Hurst found that purely stochastic phenomena present a mean value
$H \sim 0.73$ and a standard deviation $S \sim 0.09$. 

In dynamical systems, $H$ characterizes the stochastic memory in time of the process.
Hurst exponents $H> \frac12$ indicate the persistence in time of a certain trend, 
whereas exponents $H< \frac12$ indicate antipersistence, i.e. past trends tend to reverse in the future (\cite{fed88}).
An exponent $H=\frac12$ would then mean random uncorrelated behaviors with no stochastic
memory in time. 
The value $H \sim 0.73$ suggests that natural phenomena tend to present a persistent stochastic
memory in time.

Hurst's analysis has been applied in Astronomy mainly for solar phenomena. Most of the solar phenomena seem to have $H>\frac 12$, even if the actual presence of long -- memory in solar activity is still a question of debate (Oliver \& Ballester, 1998). Anyway, in the sun, there are also present some physical processes presenting different stochastic behaviors. Hanslmeier et al. (2000), showed that fluctuations of the velocity fields and the intensity fields observed in the solar upper photosphere have $H=\frac12$. 

Let's take a practical example to illustrate Hurst's R/S analysis.
The lightcurve of a CV is a discrete record in time of light intensity values that usually shows both slow and rapid stochastic luminosity variations. 
Some periodic variations are connected to the orbital motion or to the spin of the magnetic white dwarf. Besides these, CVs can also present quite a variety of irregular or quasi-periodic light oscillations. 
The shorter time scales of the flickering observed in the lightcurves  of ordinary CVs are of the order of a few seconds. However, one can also observe smooth modulations and even flares lasting hours. 
The mean luminosity $\left\langle \xi \right\rangle _{\tau }$ 
of the CV in a given time interval $\tau$ is
\begin{equation}
\left\langle \xi \right\rangle _{\tau }=\frac{1}{\tau } \sum_{t=1}^{\tau }\xi(t)
\end{equation}
where $t$ is the time coordinate. 
In the time lag $\tau$ the accumulated deviation from the mean luminosity is
\begin{equation}
X(t,\tau )=\sum_{k=1}^{t}\left[ \xi (k)-\left\langle \xi
\right\rangle _{\tau }\right]
\end{equation}
and the range $R$ of the luminosity variations
\begin{equation}
R(\tau )=\left[ \max X(t,\tau )-\min X(t,\tau )\right] _{1\leq t\leq \tau }
\end{equation}
As already said, the ratio $R/S$ follows an empirical law $R/S \sim \tau^H$ like in eq (\ref{eqh}).

Depending on the value of $H$ found, the behavior of the random fluctuations can be modeled by Wiener stochastic processes or, more specifically, by Fractional Brownian Motions (FBM). The classical Brownian motion is a particular case of FBM, with $H=\frac12$, without any stochastic memory in time. For more details, see Feder (1988).
Since FBMs possess self-similarity, they can be also easily studied with the wavelets analysis (Chui 1992, Gill \& Henriksen 1990, Simonsen et al., 1998). Using the relationships between $1/f^\beta$ noises and FBMs (\cite{low99,gao03}), there is a direct relationship between the Hurst exponent H and the parameter $\alpha$, namely $H=1-\alpha /4$.

\begin{figure}
\includegraphics[angle=0, width=8cm, keepaspectratio]{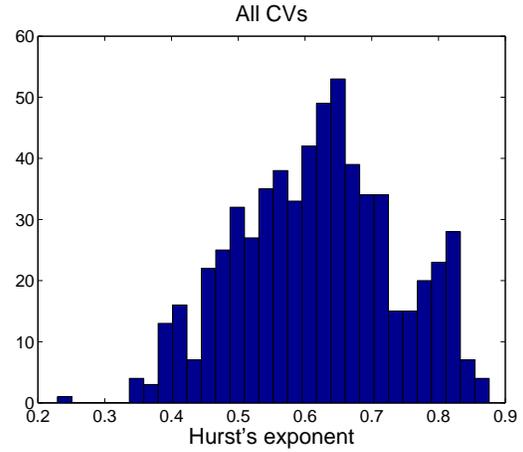}
\caption{Histogram of the Hurst exponent of all the sample of CVs.
The distribution peaks at $H\simeq0.65$. Around the value found by Hurst for most of the natural phenomena $H\sim 0.73$ there is instead a drop in the distribution.
The presence of a second bump at $H\sim 0.82$ is also possible. The distribution is non symmetric showing a tail towards the shorter values of $H$.} 
\label{fig1}
\end{figure}

In Fig. (\ref{fig1}) is shown the histogram of the Hurst exponents of all the CVs analyzed by Fritz and Bruch. We see that both persistent and antipersistent behaviors are present. We note that the peak is around $H\sim 0.65$. At the value $H\sim 0.73$ found by Hurst in most natural phenomena there is a drastic drop. The presence of a second bump at $H\sim 0.82$ is also possible. We then notice that the distribution is non symmetric with a longer tail towards the shorter values of $H$.

\section{The flickering in white light of V709 Cas}

V709 Cas (RX J0028.8+5917) is a DQ Her--type system with orbital period of 0.2225 days \cite{bonnetbidaud2001} and a $1.08M_\odot$ primary  (\cite{Ram00}) with a spin period $P_{spin}=312.8$ s (\cite{dem01}). 
We have analyzed six runs obtained in white light during a multi--site photometric campaign performed from September 22 to October 2 2000 by De Martino et al. (2002) with a three channel photometer placed at the 0.8 m Tenerife (Spain), at the 1.5 m Loiano (Italy) and at the 0.8m Beijing (China) telescopes. The integration time was in all cases 10s.

\begin{figure}
\includegraphics[angle=0, width=8cm, keepaspectratio]{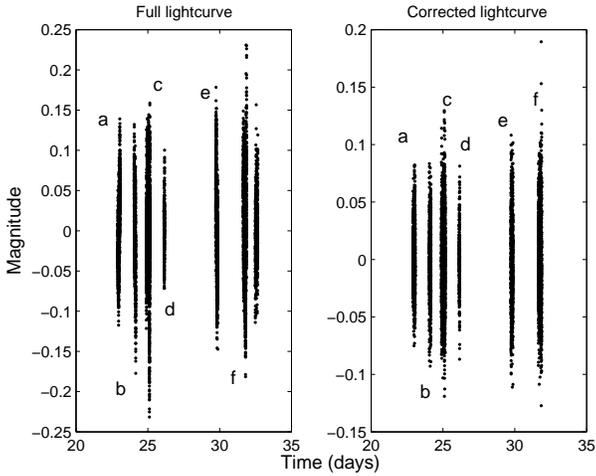}
\caption{The full set of the original and the ``corrected lightcurves'' of V709 Cas (see text). The last night of observations has been omitted in the right panel because of the presence of low quality data.}
\label{fig2}
\end{figure}

\begin{figure}
\includegraphics[angle=0, width=8cm, keepaspectratio]{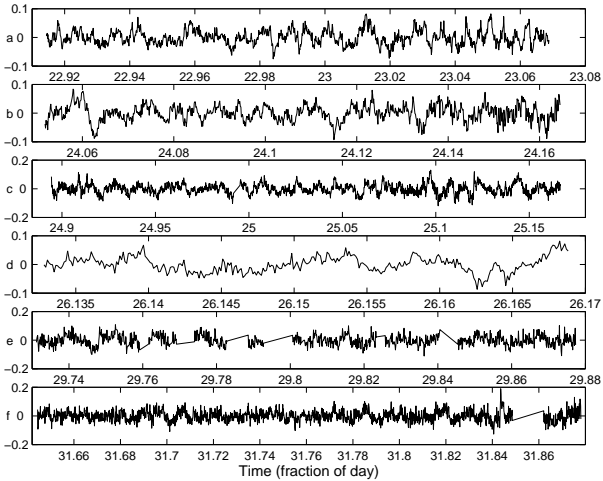}
\caption{The corrected lightcurves of the six runs (a--f) of Tab.1. 
Time is in days and magnitudes are rescaled with respect to their mean magnitude.
The discontinuities in the last two datasets are due to the presence of veils during the observations.}
\label{fig3}
\end{figure}

The lightcurves (Fig. \ref{fig2}-\ref{fig3}) show very rapid variations up to 0.2 magnitudes, indicating the presence of a strong flickering, as already reported by Kozhevnikov (2001) and De Martino et al. (2002). 
Each inset of Fig. \ref{fig3} shows the ``corrected lightcurve'', i.e. the lightcurve variations with the magnitude rescaled with respect to zero and filtered from the fluctuations due to the orbital period of the CV. The data sets are ordered as in Table 1. 
From up to down the first subplot {\bf (a)}, shows, with respect to the other subsets of data, a more active flickering, with more erratic motions, rapid and intense flares. The third and fifth subplot {\bf (c)} and {\bf (f)} show a larger variance at shorter timescales and shorter variance at larger timescales. They both present flares with larger amplitudes up to 0.2 mag. 
\begin{figure}
\includegraphics[angle=0, width=8cm, keepaspectratio]{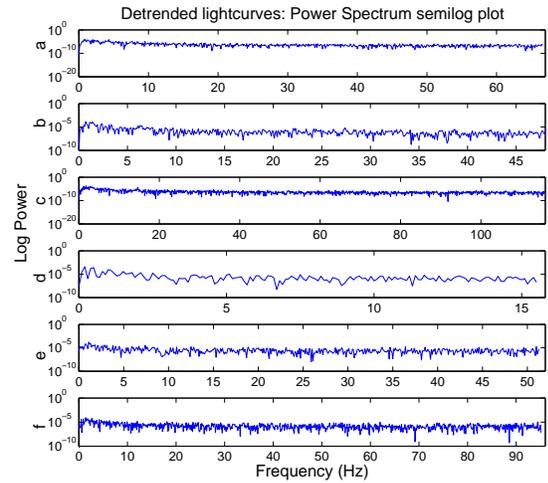}
\caption{The power spectra of the six corrected lightcurves of V709 Cas.}
\label{fig4}
\end{figure}
The power spectra of the corrected lightcurves do not show any dominating periodicity
(Fig. \ref{fig4}). The presence of some residual power ($\sim 10^{-4}$) is present at low frequencies, but it will not appreciably affect the results of our analysis.

\subsection{Data analysis}

Now we determine the parameter $\alpha$ both with wavelets and Hurst R/S analysis and plot for each run the position of V709 Cas in the parameter space $\alpha$-$\Sigma$ with the time scale resolution of 10 seconds. We used the C12 wavelets (Fritz \& Bruch, 1998). To test our software we used both wavelet-generated FBMs and $1/f^\beta$ Gaussian processes (Bak et al., 1987; Abry et al., 1995). 
Each lightcurve (a--f) was divided into series of $2^n$ bins, with $n$ is an integer varying in the interval $[1,7]$ in order to apply the prescriptions of wavelet analysis. 

Tab. \ref{tab1} gives,  for each run, the Hurst exponent $H$ and the characteristic exponent $\beta$ of the  $1/f^\beta$ process obtained with wavelets. 
The table also reports the run label, the number of data points, the maximum and 
minimum of the relative magnitudes, and the percentage difference $\Delta \%=\frac{|D_H-D_f|}{|D_H+D_f|}\times100$ between the
the fractal exponent $D$ calculated with the $1/f^\beta$ noises ($D_f=1+\frac{\beta}2$) and with the R/S analysis ($D_H=2-H$).

\begin{figure}
\includegraphics[angle=0, width=8cm, keepaspectratio]{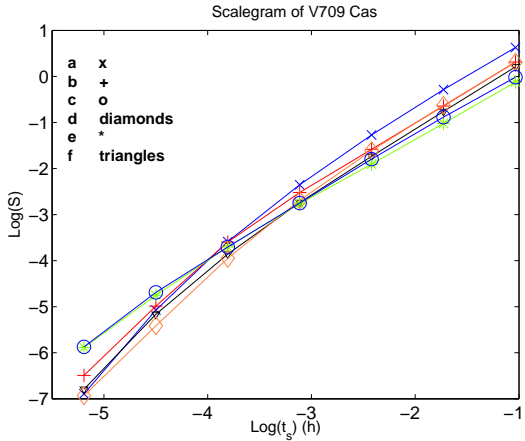}
\caption{Scalegrams of the six acquisitions of the corrected V709
Cas data. The lines corresponding to different acquisitions are
quite similar showing a stability of $\alpha$ during 10 days. 
The small differences at short time scales may be
due to different Poisson noise contributions.}
\label{fig5}
\end{figure}

The Hurst exponents of each of the six acquisitions reported in
Tab \ref{tab1} are all larger than $1/2$, indicating that the flickering 
of V709 Cas is always is a stochastic process with persistent memory in time.
The averaged value of $H$ is $0.63$ and the standard deviation is $0.05$.
The value obtained, with R/S the analysis, reported in the table, fluctuates between
$H=0.58$ and $H=0.72$. However, the differences $\Delta \%$ are rather small.  
\begin{table*}
\caption{}
\label{tab1}
\begin{tabular}{lllllll}
\hline
\hline
Run&data&Max&min&$\beta$&H&$\Delta \%$\\
\hline
a&1335&0.082&-0.075&0.82&0.59&1.2 \\
b&974&0.084&-0.093&0.72&0.64&8.3 \\
c&2325&0.130&-0.119&0.80&0.60&3.8 \\
d&312&0.081&-0.087&0.56&0.72&4.3 \\
e&1029&0.108&-0.111&0.72&0.64&4.2 \\
f&1904&0.190&-0.127&0.82&0.59&7.3
\\
\hline
\end{tabular}
\end{table*}
Fig.  \ref{fig5} presents the \textit{scalegrams} for the six corrected
lightcurves, i.e. the log-log plot of the variance $S(t_s)$ vs. the chosen time scale $t_s$.

The curves rise linearly with the time intervals and remain quite linear for all the 6 acquisitions.
This suggests that the flickering remains self--similar, with the same power--law exponent, 
from the tens of minutes down to the $10$ second time scale.
We also see that all the curves almost overlap yielding a quite stable value of $\alpha$, indicating
that, during the 10 days-period of observations, the flickering and the accretion process of the CV
remained quite stable.

\begin{figure}
\includegraphics[angle=0, width=8cm, keepaspectratio]{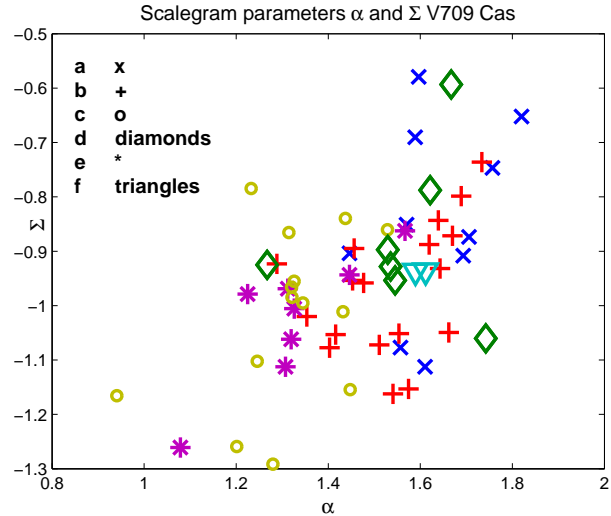}
\caption{The $\alpha$--$\Sigma$ parameter space for all the
V709 Cas data. All the values for each wavelet octave are reported. 
The relatively small region of the diagram occupied by
the data indicates a rather stable stochastic phenomenon.}
\label{fig6}
\end{figure}

Fig. \ref{fig6} shows all the values of  $\alpha$ and $\Sigma$ of V709 Cas.
The values span in the interval $0.9<\alpha < 1.8$ and $-1.3< \Sigma < -0.6$,
typical of magnetic CVs (\cite{bru92}).
In the ($\alpha, \Sigma$) plane the values obtained from V709 Cas show a trend.
$\Sigma$ is shown to decrease with decreasing $\alpha$. This would suggest that the 
persistence character of the light fluctuations and the energies implied are anticorrelated.

\section{Conclusions}
We have analyzed the flickering observed in six optical photometric runs of the IP V709 Cas.
In their pioneering work, Fritz and Bruch (1998), tried to correlate the characteristics of the flickering of CVs with their subtypes using wavelet analysis and the scalegram. The stochastic properties of flickering were described by using the two parameter space $\alpha$-$\Sigma$.
In this paper the parameter $\alpha$ was obtained both with the wavelets and the Hurst R/S analysis, the latter being a statistical robust method.
During the 10 days of the observations the stochastic properties of the flickering did not change indicating that the physics of the accretion process remained relatively stable.
We also noticed that the values of $\alpha$ and $\Sigma$ parameters, calculated at different epochs and at different time scales, show a trend suggesting that the persistence character of the light fluctuations and the energies implied in the accretion process are anticorrelated.
In the ($\alpha$, $\Sigma$) diagram V709 Cas falls in the region of magnetic systems. However, this CV was recognized as an IP only from its X-ray lightcurve while no spin modulation in the optical range is observed and the lightcurve is strongly flickering--dominated without any precise and stable dominating frequency, with the exception made for frequencies related to the orbital motion. 
Therefore, V709 Cas may represent an ideal test case where the $\alpha - \Sigma$ classification improved by the Hurst R/S analysis can be used to characterize CV subtypes, especially when the classification is uncertain.

Hurst's R/S analysis can also provide a useful tool for understanding the results of high speed photometry. In particular, it could give precious information to show down to which short time--scale the flickering can  be still considered self--similar, revealing the physics behind very short time scale phenomena. Thus, a futuristic application might be suggested for ultra-fast photometry towards the quantum limit (\cite{dra05}).

\section{Acknowledgments}

One of us, FT, also gratefully acknowledges the financial support from the 
CARIPARO Foundation inside the 2006 Program of Excellence.
DdM and AB acknowledge financial support from INAF PRIN N.17.

\end{document}